\documentclass[conference]{IEEEtran}
\IEEEoverridecommandlockouts
\usepackage{amsmath,amsfonts}
\usepackage{algorithmic}
\usepackage{graphicx}
\usepackage{textcomp}
\usepackage{xcolor}
\usepackage{siunitx} 
\usepackage{xspace}
\usepackage{pgf}
\usepackage{layouts}
\usepackage{cancel}
\usepackage[utf8]{inputenc}\DeclareUnicodeCharacter{2212}{-}
\usepackage{booktabs}
\usepackage{enumitem} 
\usepackage{textcomp}
\usepackage{mathtools}
\usepackage{subcaption}
\usepackage{amsmath}%
\usepackage{MnSymbol}%
\usepackage{wasysym}%

\def\BibTeX{{\rm B\kern-.05em{\sc i\kern-.025em b}\kern-.08em
    T\kern-.1667em\lower.7ex\hbox{E}\kern-.125emX}}

\xspaceaddexceptions{[]}
\usepackage[english]{babel}
\usepackage[linesnumbered,ruled,vlined]{algorithm2e}

\SetCommentSty{mycommfont}
\graphicspath{{figures/}}
\usepackage{amsthm}

\usepackage{fancyhdr}
\usepackage{hyperref}
\usepackage{xcolor}
\SetArgSty{textnormal}
\SetKw{extend}{extend\_and\_filter}
\SetKw{orb}{or}
\SetKw{continue}{continue}
\SetKw{addroute}{add\_route}
\SetKw{removedominated}{remove\_dominated}
\SetKw{linearclean}{linear\_clean}
\SetKw{hoppingclean}{hopping\_clean}
\SetKw{int}{int}
\SetKw{add}{add}
\SetKw{to}{to}
\SetKw{sort}{sort}
\SetKw{short}{short}
\SetKw{mergesort}{mergesort}
\SetKw{byte}{byte}
\SetKw{kwand}{and}
\SetKw{true}{true}
\SetKw{flush}{flush}
\SetKw{first}{first}
\SetKw{last}{last}
\SetKw{extend}{Extend Paths}
\SetKw{compute}{Cpt Extendable Paths}
\SetKw{succ}{succ}
\newcommand\var[1]{\texttt{#1}}

\newcommand{\ie}{\emph{i.e.,~}}
\newcommand{\eg}{\emph{e.g.,~}}
\let\oldnl\nl
\newcommand{\nonl}{\renewcommand{\nl}{\let\nl\oldnl}}
\newcommand\bestcop{\var{BEST2COP}\xspace}

\ExplSyntaxOn

\NewDocumentCommand{\makepath}{ m }
 {
 \ensuremath{
      \clist_set:Nn \mylist { #1 }
      \clist_pop:NN \mylist \listtmpvar
      \int_do_until:nNnn {\clist_count:N \mylist} = {0} {

          \listtmpvar{-}
          \clist_pop:NN \mylist \listtmpvar
      }
      \listtmpvar
  }
 }

 \NewDocumentCommand{\makesrpath}{ m }
 {
 \ensuremath{
      \clist_set:Nn \mylist { #1 }
      \clist_pop:NN \mylist \listtmpvar
      \int_do_until:nNnn {\clist_count:N \mylist} = {0} {

          \listtmpvar{|}
          \clist_pop:NN \mylist \listtmpvar
      }
      \listtmpvar
  }
 }

\ExplSyntaxOff

\newcommand{\CR}[1]{#1}

\theoremstyle{definition}
\newtheorem*{definition}{Definition}

\IEEEoverridecommandlockouts

\begin{document}

\setlength{\belowcaptionskip}{-10pt}

\title{Computing Delay-Constrained Least-Cost Paths for Segment Routing is Easier Than You Think}


\author{
\IEEEauthorblockN{Jean-Romain Luttringer\IEEEauthorrefmark{1}, %
Thomas Alfroy\IEEEauthorrefmark{1}, %
Pascal Mérindol\IEEEauthorrefmark{1}, %
Quentin Bramas\IEEEauthorrefmark{1}, %
François Clad\IEEEauthorrefmark{2}, %
Cristel Pelsser\IEEEauthorrefmark{1}}
\IEEEauthorblockA{\IEEEauthorrefmark{1} Université de Strasbourg,
\IEEEauthorrefmark{2} Cisco Systems}}

\maketitle

\begin{abstract}
With the growth of demands for quasi-instantaneous communication services such
as real-time video streaming, cloud gaming, and industry 4.0
applications, multi-constraint Traffic Engineering (TE) becomes increasingly
important. \CR{While legacy TE management planes have proven
laborious to deploy, Segment Routing (SR) drastically
eases the deployment of TE paths and thus became the most appropriate technology for many operators. The flexibility of SR sparked demands in ways 
to compute more elaborate paths.
In particular, there exists a clear need in computing and deploying Delay-Constrained Least-Cost
paths (DCLC) for real-time applications requiring both low delay and high bandwidth routes. However, most current DCLC solutions are heuristics not specifically tailored for SR.}

In this work, we
leverage both inherent limitations in the accuracy of delay measurements and an operational
constraint added by SR.
We include these characteristics in the design of \bestcop,
an exact but efficient ECMP-aware algorithm that natively solves DCLC in SR
domains. Through an extensive performance evaluation, we first
show that \bestcop scales well even in large random networks. In real networks having up to
thousands of destinations, our algorithm returns all DCLC solutions encoded as SR paths in way less than a second.
\end{abstract}

\section{Introduction}\label{sec:intro}
\CR{The fundamental challenge addressed by a routing scheme is about deploying best paths.
Internet Service Providers (ISPs) usually compute such paths according to a single additive metric, the IGP cost, which models the available resources of the network.
Generally, the IGP distance takes into account the bandwidth of each link and
is further tuned to reflect the specific needs of the ISP.
While it is sufficient for best-effort traffic, real-time flows
have strong additional requirements in terms of delay.
Our discussions with network equipment vendors indeed revealed a strong demand for ways to compute paths providing the maximal possible amount of bandwidth (to ensure a high flow quality) among the ones verifying a constraint on the end-to-end delay (the maximal latency for interactive real-time flows).
In practice, the considered delay metric is the measured propagation delay.
However, the bandwidth is not considered per se. Actually, the second considered
metric is the IGP cost. Indeed, since the latter is representative
of the bandwidth as well as the ISP's specific needs,
finding paths minimizing the IGP cost among the paths respecting the delay constraint allows the user to experience a low propagation delay and high bandwidth route while preserving the ISP resources.}


\CR{Computing such paths requires to solve the problem known as DCLC (Delay Constrained Least Cost).
This problem has attracted a lot of attention from the research
community~\cite{survey02}. Despite only considering two
(additive) metrics, DCLC is known to be NP-hard~\cite{536364}. Indeed, two
dimensions are enough to turn the total ordering existing with a single metric
into a partial one, as any path better on at least one metric compared to currently
known paths may be part of the solution and thus has to be explored. These paths are referred to as
\textit{non-dominated} paths and make up the \textit{Pareto front} of the
solution. To solve DCLC exactly, the whole Pareto front (2-dimensional because
the problem is limited to 2 metrics) must be explored. Since the latter may grow
exponentially, this family of problems is considered intractable.}

\CR{Several approximations schemes exist to solve DCLC.
However, these latter have never been deployed by operators as they do not provide sufficient guarantees in term of performance
and rely on a technology that does not scale well.
In this work, we show that DCLC can be solved exactly and efficiently without neglecting the practical deployment of constrained TE routes.
As long as two reasonable practical assumptions are verified, it is possible to build an efficient algorithm.}

\CR{Our first assumption concerns the nature of the metrics. 
Due to the apparent intractable nature of DCLC, most current solutions rely on heuristics, which are complex to deploy and do not offer strong guarantees in all cases.
However, although DCLC is computationally expensive, its complexity is often misinterpreted. Exponential
cases are unlikely to occur in practice~\cite{chen} thanks to the structures of ISP networks and the fact that their metrics lead to a limited number of distinguishable distances.
As soon as there exists few distinct values in the Pareto front, it is possible to bring strong guarantees without relying on heuristics.
Our first assumption then is that either the IGP cost or the delay metric is bounded and discrete.
While this requirement may seem strong, we argue that these metrics already meet it, the delay in particular.
Indeed, IGP distances only provide relative bounds on the Pareto front size as they ultimately depend on the ISP configurations (although
limited by the routing protocol itself). Conversely, the bounds 
provided by the delay are absolute (as they do not depend on any configurations) and often even stricter than the IGP routing limits, due to both physical limitations and the nature of delay-constrained flows.
A real-time interactive flow must meet strict guarantees
regarding its delay (at least $< 100$ms, usually closer to $10$ or even $2$ms).
In addition, even though the precision in memory of the measurements can be
high, \ie nanosecond, no delay measurement technique can claim to be that accurate in worst-case scenarios. ISPs
are aware that measured delays are an estimation, not a guarantee. These delays (usually measured with OWAMP~\cite{RFC4656}/TWAMP~\cite{RFC5357}) have a limited \textit{trueness} (as
defined in ISO 5725-1~\cite{ISO5725}) or accuracy, even with efficient hardware
PTP time stamping systems or accurate two-way delay estimators. Due to both its
inherent variation depending on physical properties, clock synchronization or inconstant packet processing delays~\cite{RFC7679,
RFC2681}, we argue that truly distinguishing the delay of a path at the
granularity of the micro-second is difficult if not impossible nowadays.
Even the finest delay estimation is bounded and discrete by essence,
allowing us to predict the size of the Pareto front and control the complexity of our algorithm.}

\CR{
Our second assumption concerns the technology at play.
Current solutions do not usually consider the
deployment of the computed paths. Consequently, they rely on RSVP-TE which does not impose 
any additional constraints on the paths to deploy. However, they thus scale poorly as RSVP-TE suffers from well-known scalability issues since both the number of control-plane messages and of forwarding states scale with the quantity of TE paths~\cite{7417124}.
We thus design our algorithm for a specific technology, Segment Routing~\cite{7417124} (SR). SR is the new state-of-the-art TE and fast-reroute technology now deployed in most ISPs. SR relies on a very lightweight control-plane as forwarding states are carried within the packets.
More precisely, SR implements source-routing by translating forwarding paths into lists of
segments (routing instructions). This
list is then encoded within each packet and used by routers to forward said packet. However, the size of this list is limited, as only
\textit{SEGMAX}$~\approx10$ segments may be imposed on each packet at line-rate
with today's best hardware. If handled naively, this new constraint
can considerably increase the problem's complexity, as it may seem necessary to explore a now three-dimensional Pareto front. We design our algorithm with SR in mind by exploring the solution space in a way that
leverages the SEGMAX constraint. We are thus able to efficiently deploy segment lists that respect both a constraint on their sizes and their delay while minimizing the IGP distances.}

In summary, we leverage the two aforementioned properties (the inaccuracy of delay measurements and the SEGMAX constraint) to
provide a straightforward, efficient algorithm suited for practical TE usage: \bestcop (\textbf{Best Exact Segment Track for 2-Constrained Optimal Paths}).
\bestcop is, to the best of our knowledge, the first algorithm able
to solve DCLC efficiently in any scenario for large SR domains, and thus the first deployable DCLC scheme ever.
In short,
it computes all the paths verifying two constraints (delay and number of
segments) while optimizing the IGP distances. \bestcop can provide DCLC solutions for large realistic
networks in a time period acceptable for the routing convergence, \ie way less
than a second.

In Sec.~\ref{sec:ps}, we formally define the \textbf{2COP problem} for
solving DCLC in an SR domain. Sec.~\ref{sec:algo} sketches \bestcop before we evaluate its performance in
Sec.~\ref{sec:eval}. We conclude the paper discussing the related work in
Sec.~\ref{sec:rl} and summarizing our achievements and future works in
Sec.~\ref{sec:conc}.

\section{2COP, or Solving DCLC within a SR Domain}\label{sec:ps}

In this section, we formally introduce and define all notations and concepts used to design \bestcop.
More precisely, we detail the problem we aim to solve, the construct we use to encompass Segment Routing natively, and how it is used to our advantage along with the delay characteristics.

\subsection{Problem Statement, or the Need for an SR Graph}\label{ssec:ps}

We aim to solve an SR variant of the DCLC problem, considering the IGP cost, the propagation delay, and the number of segments.
We refer to this problem as 2COP. For readability purposes, we denote:
\begin{itemize}
  \item M$_0$ the metric referring to the number of segments, with the constraint $c_0 = \mathit{SEGMAX}$;
  \item M$_1$ the delay metric, with an arbitrary constraint c$_1$;
  \item M$_2$ the IGP metric being optimized.
\end{itemize}
We also rely on these generic notations to highlight that the problem remains the same for any couple of metrics. Besides, the problem keeps the same complexity even if M$_2$ is also a constrained metric.

\definition{}
\textit{2-Constrained Optimal Paths (2COP)}:
Given a source $s$, 2COP consists in finding, for \textit{all} destinations, a \textit{segment list} verifying
two constraints, c$_0$ and c$_1$, on
the number of segments (M$_0$) and the delay (M$_1$) respectively,
while optimizing the IGP distance (M$_2$). We denote this problem 2COP(s, c$_0$, c$_1$).
\hfill$\blacksquare$

\bestcop solves 2COP by looking for all feasible distances (\ie satisfying c$_0$
and c$_1$) optimizing M$_2$. Note that 2COP is distinct from DCLC because of
c$_0$. For example, let us refer to Fig.~\ref{fig:exampleraw} and consider DCLC
for $c_1 = 7$ from node s to p. The solution is the path $\makepath{{(s_1,n)}, {(n_1,o)},
{(o,p)}}$ having an IGP cost of 4 and a delay of 6.49. However, the latter is not a solution of 2COP(s, 2, 7) towards p. Indeed, we will see that this path translates to three segments and violates c$_0 = 2$.

SR implements source
routing by pre-pending packets with a stack of segments.
Segments can be seen as a list of checkpoints the packet has to go through sequentially, be it a node or a specific
interface. SR mainly uses two types of segments: \textit{node segments} and
\textit{adjacency segments}. A node segment specifies a node as a
checkpoint. A node segment representing a destination $v$ is interpreted by a
router as \textit{forward the packet to $v$ (through the best IGP path(s))}.
Since SR enables ECMP by design, flows are load-balanced among best
paths to $v$. On the contrary, an adjacency segment indicates that the router has to forward the packet through a specific local interface.

Thus, to encompass SR natively while solving DCLC, we rely
on a structure for which the IGP costs, delays, and the number (and type) of segments used to build the
segment lists are direct and natural properties.
We call this structure an \textit{SR graph}. An SR graph represents the segments
as edges, whose weights are the (M$_1$ ; M$_2$) distances of the underlying path
or adjacency encoded by the segment.

Exploring paths on the SR graph is equivalent to exploring stacks of segments
and the paths they encode. A path requiring $x$ segments is represented as a
path of $x$ edges in the SR graph (agnostically to its actual length in the raw
graph). Thus, within an SR graph, one can simply check that $x < c_0$ to verify
the constraint on M$_0$.
\CR{In an SR domain, the SR graph is computed by default for any TE usages including fast-reroute.
In our case, only some extra information needs to be added in order to correctly handle multiple metrics, which does not generate a significant overhead to the SR graph computation.}




\subsection{The SR Graph Construction}

\begin{figure}
  \centering
  \includegraphics[scale=0.65]{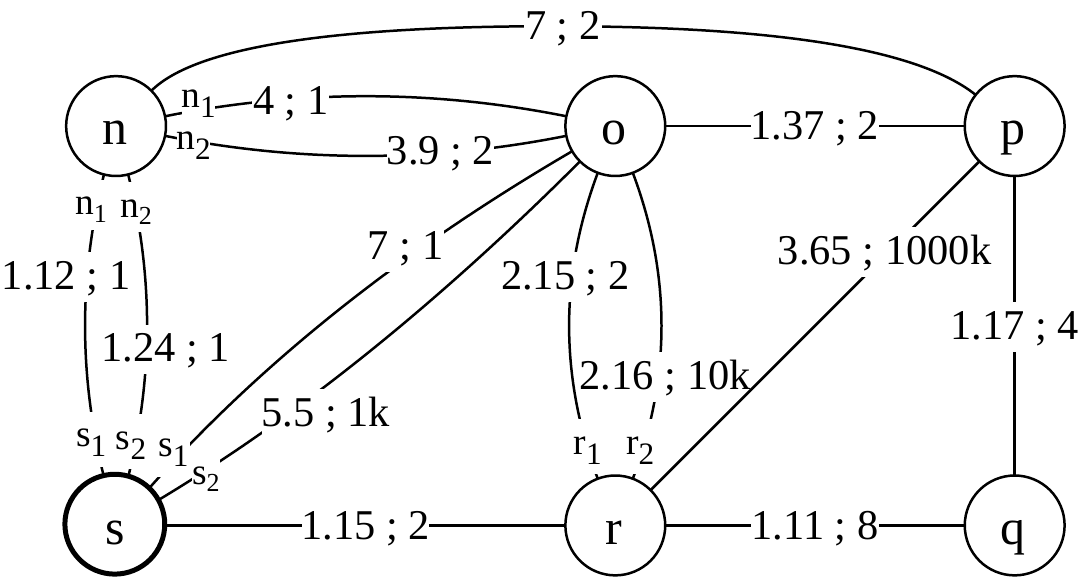}
  \caption{The raw network graph $G = (V,E)$ is a multigraph, with weighted edges.
  The weight of the edges is represented as a couple (delay;IGP cost). While there
  sometimes only exist a single edges between two nodes, such as $(s,r)$, we otherwise distinguish between parallel nodes such as $(s_1,n)$ and $(s_2, n)$.}
  \label{fig:exampleraw}
\end{figure}

\begin{figure}
  \centering
  \includegraphics[scale=0.65]{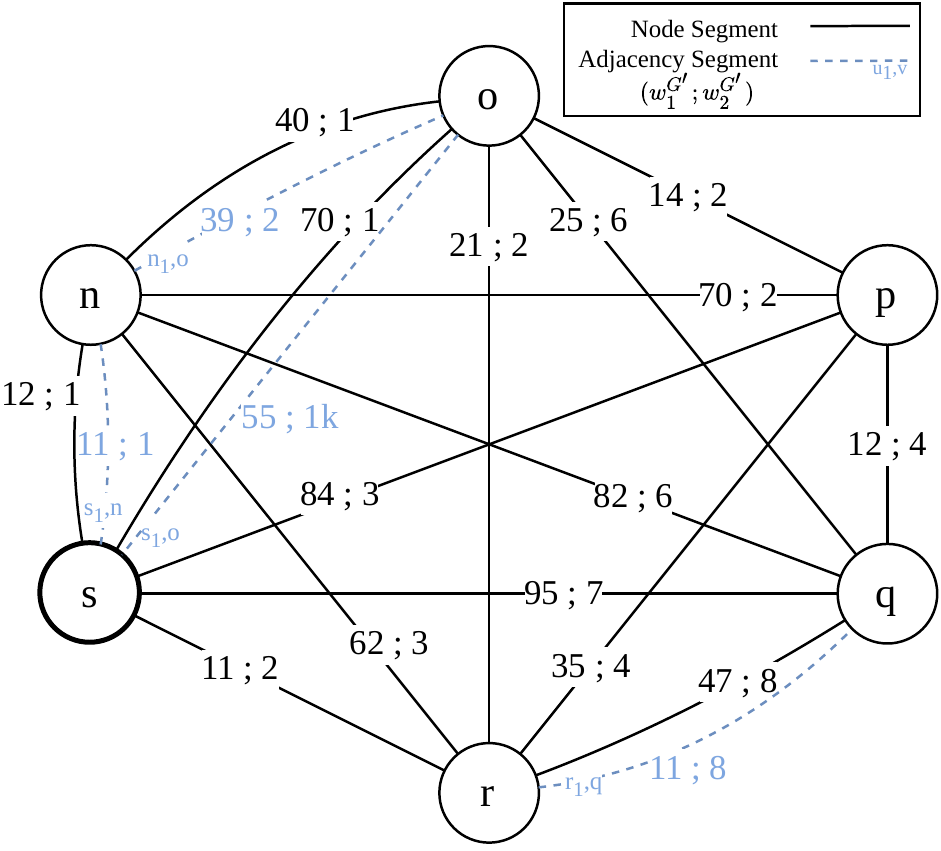}
  \caption{The SR graph $G'(V,E')$ encodes segments as edges. Plain black edges represent node segments, \ie, sets of ECMP best path, while dashed blue edges represent link in the original graph $G$, making $G'$ a full-mesh at least. Adjacency segments, \eg $(s_1,n)$,
  are only represented if they are not dominated by other segments.}
  \label{fig:examplesr}
\end{figure}

Throughout this section, we use Fig.~\ref{fig:exampleraw}
and~\ref{fig:examplesr} to illustrate the SR graph construction. While the
former provides an arbitrary raw graph, the later gives its resulting SR
counterpart. We start by describing the notations we use, in particular
regarding multigraphs, as both the SR and raw graphs fall in this category.

Let $G = (V, E)$ be the original graph, where $V$ and $E$ respectively refer to
the set of vertices and edges. As $G$ can have multiple parallel links between a
pair of nodes $(u,v)$, we use $E(u,v)$ to denote all the direct links between
nodes $u$ and $v$. When necessary, we denote a specific link $(u_x, v)$, $x$
specifying the considered interface (a local number to $u$). Each link possesses two
weights: its delay and its IGP cost. The delay and the IGP cost being additive
metrics, the M$_1$ and M$_2$ distances $(d_1^G ~;~ d_2^G)$ of a path are simply
the sum of the weights of its edges.


From $G$, we create a transformed multigraph, the SR graph denoted $G' = (V,
E')$. While the set of nodes in $G'$ is the same as in $G$, the set of edges
differs. Indeed, $G'$ encodes segments as edges representing either adjacency
segments (blue dashed edges encoding some adjacencies of $G$) or node segments,
encoding sets of best ECMP paths. For instance, in Fig.~\ref{fig:exampleraw},
there exist three paths in $G$ from $n$ to $r$ which have an optimal IGP-cost of
3: $\makepath{{(n_1,o)} , {(o,r)}}$ with distance $(6.2 ; 3)$,
$\makepath{({n_1,s)} ,{(s,r)}}$ with distance $(2.3 ; 3)$ and
$\makepath{{(n_2,s)}, {(s,r)}}$ with distance $(2.4 ; 3)$. They are thus all
represented by the unique link $(n,r)_{G'}$ (in plain black in $G'$). When using
a node segment specifying the destination $r$ from $n$, the traffic is load
balanced across the three paths.

Note first that we do not use $6.2$ as the delay value in $G'$ but $62$. Indeed, while
the delay can be represented as a precise floating number, its actual accuracy
is limited. We can safely round the M$_1$-weights (delays) without losing
relevant discriminating information, reducing so the complexity of 2COP as we will detail in the following. Thus the path from $n$ to $r$, relying on
the best IGP distance (a node segment in $G'$), has distances $(62 ; 3) =
(40+22; 1+2)$, where 22 is the delay 2.15ms times 10 rounded up at the 0.05
accuracy grain. Second, we chose $62$ in particular as the delay of the node segment because the only delay
guarantee of a node segment is to not exceed the worst delay of all its ECMP
paths.

In summary, a node segment encoding the whole set $P_G(u,v)$ of ECMP best paths
between two nodes $u$ and $v$ is represented by exactly one edge in $E'(u,v)$.
Its M$_2$-weight, $w_2^{G'}((u,v))$, being the common M$_2$-distance of $P_G(u,v)$, its M$_1$-weight, $w_1^{G'}((u,v))$, is defined as the maximum M$_1$-distance among all the paths in $P_G(u,v)$.
Thus, links representing node segments in $G'$ verify the following:
\[
\begin{array}{l}
    w_1^{G'}((u,v)) = \max_{P\in P_G(u,v)} d_1^G(P)\\
    w_2^{G'}((u,v)) = d_2^G(P)\quad \text{for any }P\in P_G(u,v)
\end{array}
\]

In addition to this unique node segment, $E'(u,v)$ may contain \emph{adjacency
segments} (dashed blue edges in Fig.~\ref{fig:examplesr}) to force the packet to
go through a specific interface. An adjacency segment corresponds to an
edge in the graph $G$ and is represented by an edge $(u_x, v)$ in $E'(u,v)$
whose M$_1$-weight, resp. M$_2$-weight, is exactly the M$_1$-weight, resp.
M$_2$-weight, of the corresponding link in $G$. Note that if the node segment
between two nodes has both a better delay and a better cost than any direct link
between them, there is no point in using an adjacency segment. More generally
speaking, an adjacency segment exists in $G'$ only if it is not dominated by
other segments. Formally, an adjacency segment is represented by an edge $(u_x,
v)_{G'}$ if it is not dominated by the node segment $(u,v)_{G'}$, \ie if
$d_1^G((u,v)) > w_1^{G'}((u_x,v))$, or by any other non-dominated adjacency
segments numbered $y$, $(u_y,x)$, \ie if $w_1^G((u_y,v)) > w_1^{G'}((u_x,v))$ or
$w_2^G((u_y,v)) > w_2^{G'}((u_x,v))$. For example, in Fig.~\ref{fig:exampleraw},
the best path from $n$ to $o$ has a distance of $(4; 1)$ and is translated to
the node segment as a link with same values in Fig.~\ref{fig:examplesr}. Since
there exists another direct link between both nodes with a lower delay, $(3.9 ;
2)$, we add an edge $(n_1, o)$ with
distances (39 ; 2) to $G'$. One can then force the corresponding adjacency
segment to save delay.

In practice, the SR graph $G'$ can be built for all sources and destinations
thanks to any APSP algorithm to compute the weights of each node segment in $G'$. We consider this construction as a shared input for \bestcop, as this
transformation is inherent to SR and applies network-wide. \CR{Note that this
computation is unlikely to be performed by the router itself, but rather by a Path Computation Element~\cite{RFC4655}, which may be located within a controller. The overhead added to this construction by our specific transformation is
negligible; it consists in the addition of the delay information, in particular to select non-dominated (adjacency) segments.}

Thanks to this
specific construct, we can now illustrate the sets of paths we want to retrieve
when solving 2COP. We said in Sec.~\ref{ssec:ps} that while path
$\makepath{{(s_1, n)}, {(n_1, o)} , {(o, p)}}$, having a distance of (6.49 ; 4),
solves DCLC for c$_2$ = 7 and for destination $p$, it does not solve 2COP(s, 2,
7). Indeed, we can now clearly see in $G'$ (Fig.~\ref{fig:examplesr}) that to
achieve this path, 3 segments are required: $\makesrpath{{(s_1,n)}, {(n,o)},
{(o,p)}}$, which makes it non SR-feasible with a segment budget of 2. The
solution to 2COP(s, 2, 7) is a list of 2 segments (\ie a path of 2 edges in
$G'$): $\makesrpath{{(s,r)}, {(r,p)}}$, encoding the path $\makepath{{(s,r)},
{(r_1, o)}, {(o, p)}}$ in $G$. Note that this physical path has a distance of
(4.6 ; 6).

\subsection{An SR Graph with True Measured Delays}

In this section, we explain how the characteristics of real ISP networks are used to our advantage
and translate in the construct we have detailed.

DCLC is pseudo-polynomial~\cite{gareycomputers}.
More precisely, it is polynomial in the smallest largest weight of
the two metrics M$_1$ and M$_2$ (once translated to integers).
As long as one of the metrics possesses only a limited number of distinct values, the problem is tractable and can be solved efficiently, since the limited range of the metric restricts the number of non-dominated distances.
The metric (and so, the number of distinct distances a path can have) can be bounded and its accuracy coarse by nature, or c$_1$ can be small enough to sufficiently reduce this number of values.
Although our solution can be adapted to fit any metric, we argue that M$_1$, the propagation delay, is the best candidate
and will, most of the time, have the lowest number of distinct values.

The delay is usually constrained through a strict bound (always lower than 100ms in practical cases).
In addition, while the delay of a path is generally represented by a precise number in memory, the actual accuracy of the measured delay of an edge in $G$ is far lower.
Indeed, due to the inherent lack of accuracy of any delay measurements, discriminating paths having less than a 0.1ms difference can be challenging if not impossible in the worst conditions. In that case, floating numbers representing the delays can be rounded to integer taking 0.1ms as unit.
Since the delay is also bounded through its constraints, the number of distinct, discriminable delay values is likely to be very limited.
This allows us to easily bound the number of possible non-dominated distances to $c_1 \times t$, with $t$ being the level of accuracy of M$_1$ (the inverse of the delay grain). For example, with $c_1 = 10$ (in ms) and a delay grain of 0.01 ms, we have $t = \frac{1}{0.01} = 100$ and so only $1000$ distinct (rounded) values to manipulate with \bestcop.


In practice, this numerical value is controllable for
solving 2COP even if c$_1$ is not a strict constraint. Indeed, let us recall that we can also leverage the limited
number of segments, c$_0$ := SEGMAX. We are limited to c$_0 \approx 10$ segments, \ie paths
of 10 edges in the SR graph $G'$. Regardless of the constraint $c_1$, we know
that a feasible SR path will not exceed an M$_1$-distance of the maximum
w$_1^{G'}$ weight on the SR graph times c$_0$. If we denote by $\mathcal{S} \times t$
the maximum edge delay in $G'$ -- once rounded to integer with an accuracy level of $t$ -- we know that
a feasible SR path has a delay of at most $10 \times \mathcal{S} \times t$.
In any cases, the number of possible distinct M$_1$-weight of SR-feasible paths in $G'$ is bounded by $\Gamma = \min(c_1, c_0 \times \mathcal{S}) \times t$.

For a rounded delay, it then becomes sufficient to store only the best M$_2$-distance (indexed on its respective M$_1$-distance), leading to a Pareto front that can be stored in a static array of size $\Gamma$.
In other words, there are at most $\Gamma$ non-dominated pairs of distances to be stored and 2COP is polynomial in $\Gamma$.
The complexity of 2COP is thus controllable.
With a small enough delay constraint, the level of accuracy $t$ of the delay can be increased and 2COP solutions can remain exact since $1/t$ becomes smaller than
the inherent delay measurement error.
Keeping constant the constraint-accuracy product makes the error margin
constant relatively to c$_1$.
As an example, maintaining $\Gamma = 1000$ with c$_1$=100ms results in
$t=10$ and in an error margin of 0.1ms. With c$_1$=10ms, the accuracy level $t$ can be increased to $100$, resulting in an error margin of $0.01$ms. In all cases, the error margin is 0.1\% relative to c$_1$.
If c$_1$ is a loose constraint, keeping $\Gamma = 1000$ results in weaker approximations
as $t$ has to decrease, making the error margin become greater than the measurement inaccuracy. However, it allows the problem to remain tractable.

While the number of paths to manipulate in $G'$ is limited thanks to the aforementioned properties, it may be still considerable when $V$ increases.
Fortunately, we can once again leverage SEGMAX to cut down the exploration space.
Since we are only interested in SR-feasible paths, we do not need to explore paths requiring more than c$_0$ segments.
Using a variant of the Bellman-Ford algorithm on the SR graph, this can be done easily as the path exploration naturally iterates over the number of segments, thanks to both the algorithm's design and the SR graph representing segments as edges.
\bestcop visits $G'$ paths of $i$ segments at its $i^{th}$ iteration, allowing us to stop after very few iterations at worst (all paths discovered afterwards
exceed c$_0$).

\section{The BEST2COP Algorithm}\label{sec:algo}
\begin{figure*}[!ht]
    \centering
    \includegraphics[scale=0.9]{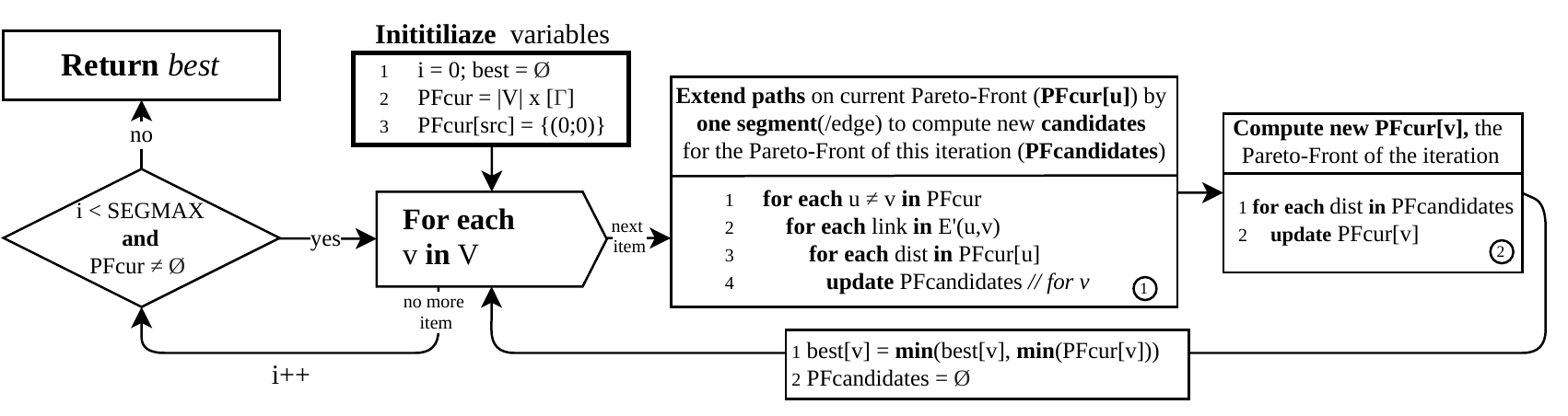}
    \caption{\bestcop algorithm. \bestcop works by exploring paths of increasing length on $G'$. Non-dominated paths are extending by one edge. The algorithm ends at the SEGMAX$^{th}$ iteration or when progress stops.}
    \label{fig:BEST2COP}
\end{figure*}

In this section we describe \bestcop, our algorithm efficiently solving 2COP by
leveraging properties formalized in the previous section.
\CR{We propose here a high-level description, but the interested reader can find its implementation online\footnote{https://github.com/talfroy/BEST2COP}.
While the implementation is designed for high performances, we omit here several details regarding its precise data structures (even though these latter play an
important role in \bestcop's overall efficiency).}
\CR{Like the SR graph computation, \bestcop can be run on a centralized controller but can also be directly launched by each router.}

Simply put, at each iteration, \bestcop starts by extending the known paths for each node by
one segment (\ie one edge on the SR graph) in a Bellman-Ford
fashion (a not-in-place version to be accurate); at the main difference that we
consider here all non-dominated paths, \ie the Pareto front. Second, newly found extended paths are filtered to reflect the new Pareto front.
The remaining one will then be extended themselves, but not before the next
iteration. Thanks to SEGMAX, these two steps only need to be performed $\approx
10$ times. Indeed, since we explore paths segment by segment, paths of $i$
segments (\ie $i$ edges in the SR graph) are explored at iteration $i$. All paths not explored before the tenth iteration require more than SEGMAX
segments and can be ignored.

The good performance of \bestcop does not only result from a cut in the
exploration space, but also from well-chosen data structures.
Since the limited accuracy of the measurements bounds
the number of non-dominated distances to $\Gamma$ at
each step, we can manipulate arrays of fixed size.

Fig.~\ref{fig:BEST2COP} sketches the main steps of \bestcop.
We focus on the \textit{PFcur} structure
for the sake of simplicity.
The number of elements within PFcur is bounded
by $\Gamma$. PFcur stores, for the
current iteration and each vertex, the Pareto front of the distances indexed on
their delay (M$_1$). Since \bestcop explores paths segment by segment, PFcur
will contain, at the $i^{th}$ iteration, all distances within the Pareto front
encodable in exactly $i$ segments. In particular, \bestcop only needs to store in PFcur the best M$_2$-distance for a given M$_1$ index, as
we aim to find least-cost paths.



In the initialization, we set that the only known best distance is the distance
to the source $src$ itself, $(0,0)$ (\ie PFcur[src][0] = 0). At iteration $i$ (and as shown in box 1), for
all predecessors $u$ of each node $v$ (\ie potentially all $u$ in $V$ since
$G'$ is a full mesh), \bestcop extends all the non-dominated distances to $u$
(PFcur[u]) discovered at the previous iteration $i-1$ by all weights in
$E'(u,v)$ (box 1, Line 2). By combining all distances to $u$ discovered at
iteration $i-1$ with all weights of parallel links linking $u$ to $v$ in $E'$, we
compute \textit{candidate} distances of $i$ segments towards $v$. Note that since
M$_2$-distance towards $v$ are indexed on M$_1$, if a newly discovered
M$_2$-distance is worse than the one currently sitting at the same M1-index, it
means that the distance is dominated and that there is no point in keeping it
(in box 1, it is basically the test performed in the update function of line 4).
However, these new distances to $v$ may not be on the Pareto front as they can
be dominated by other values newly discovered, stored in other indexes. We thus refer to them as
\textit{PFcandidates}. From the set of candidate paths, we still need to extract
the new Pareto front of the current iteration (Box 2), which is stored in
PFcur[v] and will in turn be extended at the next iteration. This is the
purpose of the second update function that checks whether the candidate is
actually non-dominated.
For a given node, at the end of an iteration, the best known distances to it
were correctly updated and will serve as a basis for the next iteration. We can
then also safely record the current best path that minimizes the cost and
respects by design the constraint c$_0$ (using the array denoted
\textit{best} in the flow chart).

In reality, \bestcop is far more versatile and able to optimize any of the three metrics (M$_0$, M$_1$ or M$_2$) with possibly constraints on all metrics.
For example, referring back to Fig.~\ref{fig:examplesr}, \bestcop is able to
return the solution to 2COP(s, 3, 70) towards p
optimizing M$_0$ or M$_1$ (\makesrpath{{(s,r)}, {(r,p)}}) or optimizing M$_2$ (\makesrpath{{(s_1,n)}, {(n_1, o)}, {(o,p)}}). In addition, with only few adjustments on the returned structured, \bestcop is able to return, upon a single run, all non-dominated distances respecting up to three constraints (c$_0$, c$_1$ and an additional c$_2$ in the most general case). Thus, if one decides to use a stricter c$_1$ constraint, \eg 2COP(s, 3, 65), the new constrained path (\makesrpath{{(s_1,n)},{(n_2, o)}, {(o,p)} }) can already be found within the returned structure.

The time complexity of \bestcop is showcased in the flowchart. For
the $|V|$ possible neighbors of $|V|$ nodes, we extend up to $\Gamma$ non-dominated distances by the $L$ direct parallel links between them. This
procedure is repeated up to SEGMAX times, leading to a time complexity of
$O(\mathit{SEGMAX} \times |V|^2 \times L \times \Gamma)$.

For the performance evaluations, we will consider that:
\begin{itemize}
 \item $\mathit{SEGMAX} = 10$ as it matches current hardware capacity;
 \item $L = 2$: on average, in $G'$, one can expect that the total number of links in $E'$ is lower than $2|V|^2$.
 Indeed, adjacency segments are not likely to be numerous within transformed graphs, as they tend to be dominated;
 \item $\Gamma = 1000$: while controllable to reflect the expected product trueness-constraint on M$_1$, we consider an accuracy level $t$ of 10 (0.1ms accuracy) regarding a maximal constraint $c_1 =$ 100ms.
\end{itemize}






\section{Performance Evaluation}\label{sec:eval}


This section evaluates the computing time performance of \bestcop.
\CR{We focus here solely on \bestcop's performances rather than relying on a comparison.
As mentioned in Section~\ref{sec:intro}, most solutions rely on heuristics resulting in a poor exploration of the solution space in worst-cases. Conversely to these methods, \bestcop provides controllable results and very good performance impervious to peculiar worst-cases. Furthermore, no existing schemes are specifically designed for SR.
Upgrading them to handle SR raises several challenges, as minimal modifications would drastically increase their execution times. Such a fair comparison is left for future work.}

First, it is
worth to notice that without any graph-based assumptions except the ones
mentioned above (\ie just setting $|V| = \Gamma = 1000$ and with an average of
two parallel links per connection, $L = 2$), \bestcop never takes more than one
minute to explore its full iteration space. That is, when \bestcop is forced to performs its
maximum number of operations on any graph having these characteristics,
solving 2COP \textit{cannot} exceed one minute. This extreme upper bound is far from \bestcop's real
performance, as its data structures were virtually filled up to push
it to its limits. In practice, when considering concrete underlying networks,
even random ones, \bestcop can easily deal with average or worst-cases in less
than half a second.


Conversely to the vast majority of existing evaluations related to TE routing
algorithms, we focus on challenging scenarios implying large networks.
Moreover, we do not rely on a strict delay constraint to simplify the problem. First, we only consider the
largest one that is practically relevant to DCLC, c$_1 < 100$ms. Second,
we do not ignore distances whose pruning in the SR graph can reduce the
exploration space, which stresses our
solution as much as possible. Formally, our evaluations are designed such that
\bestcop returns, for all $n \in V$, the whole $2COP(s,10,100)$ set.

Given the difficulty to find real or inferred graphs having two valuation
functions, we leverage the characteristics of SR Graphs (namely, their
fully-meshed structures) to generate numerous scenarios. We nevertheless
conclude on evaluations performed on real ISPs with real IGP costs and delays.
For all the evaluations, we rely on a 4,2 GHz Intel Core i7 CPU.
\CR{While parallelizing \bestcop through slight tweaking is possible, this additional evaluation is left for future work. We show here only the results of a purely sequential approach.}

\subsection{SR Graph with Random Valuation}\label{sec:2fm}

An SR Graph $G'$ is at least a full-mesh when the original
graph $G$ is connected. We use this
convenient property to ease the generation of SR graphs for our evaluations.

We generate complete graphs of $|V|$ nodes having $|E'| = 2|V|^2$ edges,
creating so a \textit{double full-mesh} graph. One systematic additional link is
enough to mimic unfavorable practical cases, as realistic topologies tend to
possess a low average number of adjacency segments once converted. Regarding IGP
weights, we chose them uniformly at random between $1$ and $2^{24}$ (the
maximum possible IGP cost with current IGPs). Propagation delays are
uniformly distributed at random between $1$ and $\mathcal{S} = \{100, 500, 1000\}$.

\begin{figure}
  \centering
  \includegraphics[scale=0.6]{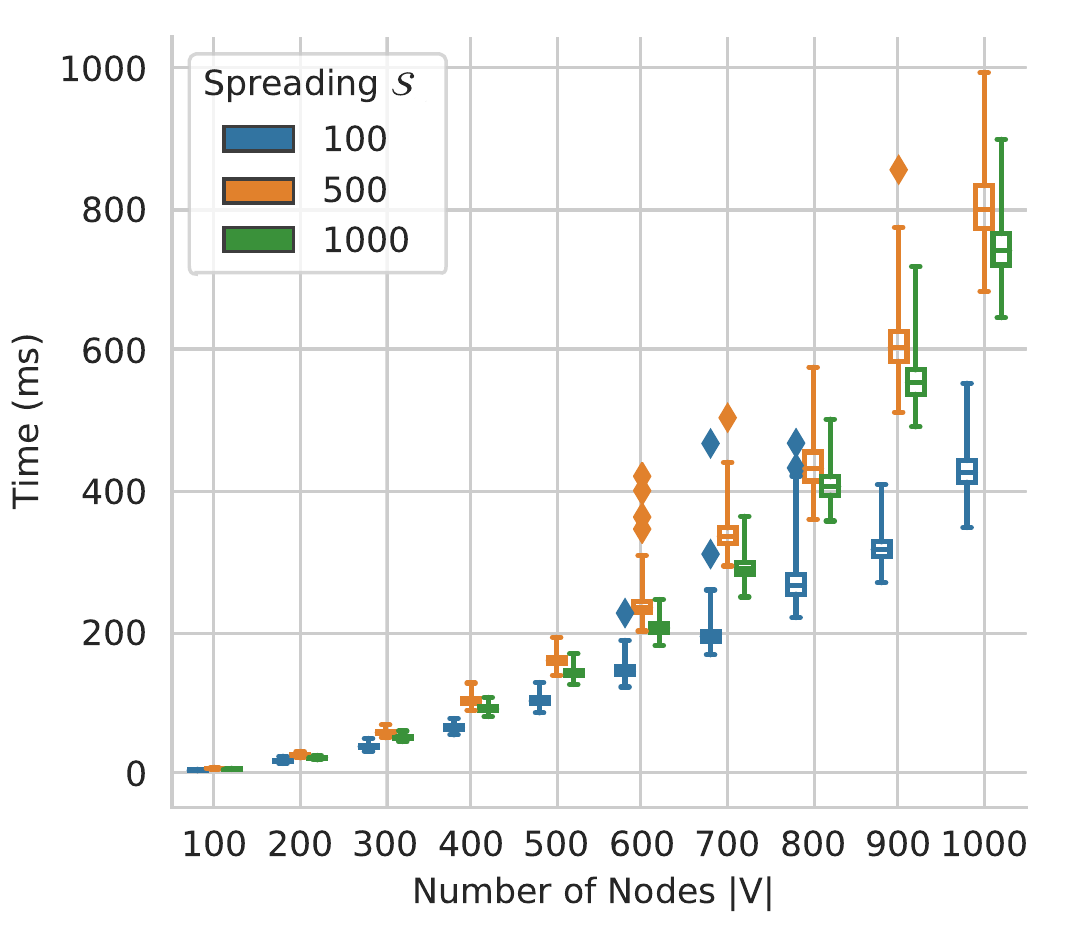}
  \caption{\bestcop worst-case when considering randomly weighted SR graphs with three spreading valuations.
  \bestcop remains always under the second and scales decently regarding $|V|$.}
  \label{fig:eval:randfm}
\end{figure}

Since we set $\Gamma$ at $1000$, picking delay weights higher than $1000$ (with a higher delay spreading) is too advantageous by design
as many distances will exceed the constraint. We perform these tests for $|V|$
ranging from $100$ to $1000$ (with steps of 100). To account for the randomness
of both valuation functions, we generate, for each $|V|$, 30 differently
weighted distinct topologies, and run \bestcop on $|V| \times 0.1$ nodes as
sources. This evaluation is not advantageous as we do not benefit from any
pattern generated by realistic networks. The resulting computing times are
shown in Fig.~\ref{fig:eval:randfm}.

This random weights evaluation exhibits the efficiency of \bestcop: its
execution time stays under one second in all of its runs.
\bestcop scales well enough with the dimension of the network which is the
critical performance parameter (quadratic in $|V|$). It is also worth noticing
that a spreading value of 500 leads to the worst time results (label
$\mathcal{S} = 500$), while a value of 1000 or only 100 leads to
a slightly better or a very notable decrease in execution time respectively.

The M$_2$ distance spreading has indeed a great impact on the filling rate of
our data structures as it can mitigate the growth of the Pareto front.
When $\mathcal{S} = 100$, which is the best-case scenario shown in
Fig.~\ref{fig:eval:randfm}, the first iterations of \bestcop have a Pareto front
size bounded by only $i \times 100 \leq \Gamma$. With larger spreading values
(and so weights), the full distance spreading regarding $\Gamma$ comes faster
(\ie with a smaller $i$) but only to some extent. This means that large
spreadings can also be advantageous: many paths within the network are bound to
have a delay higher than 1000. The number of ignored paths thus increases
significantly because many distances become greater than the constraint c$_1$.
Since there is no need to store them, \bestcop can ignore many paths and thus end up with a very fast execution time.

We have shown that \bestcop performs well with random weighted SR
graphs even when valuation bounds are not favorable. However, we considered here SR graphs that were not constructed through
the translation of an existing raw graph $G$. In the next section, \bestcop will
benefit from real raw networks' structures and valuations. SR graphs translated from real topologies are likely to
be vastly simpler, with more sparse and possibly aligned valuations.

\subsection{More Realistic Scenarios}\label{ssec:mrs}

The performance of \bestcop being already promising on non-advantageous
scenarios, we now analyze its performance in realistic cases.
Our basic settings are left unchanged, \ie $\Gamma =
1000$ and \bestcop still does not take advantage of any distance pruning to
reduce $G'$. We first evaluate \bestcop's execution times on a large network
topology with real IGP weights but random delays. Then, we consider real
but smaller network structures having both real IGP weights and delays. Our
goal is to show at which extent \bestcop can benefit from concrete network
characteristics, making it efficient enough to be deployable for real-life
cases.

\begin{figure}
  \centering
  \includegraphics[scale=0.5]{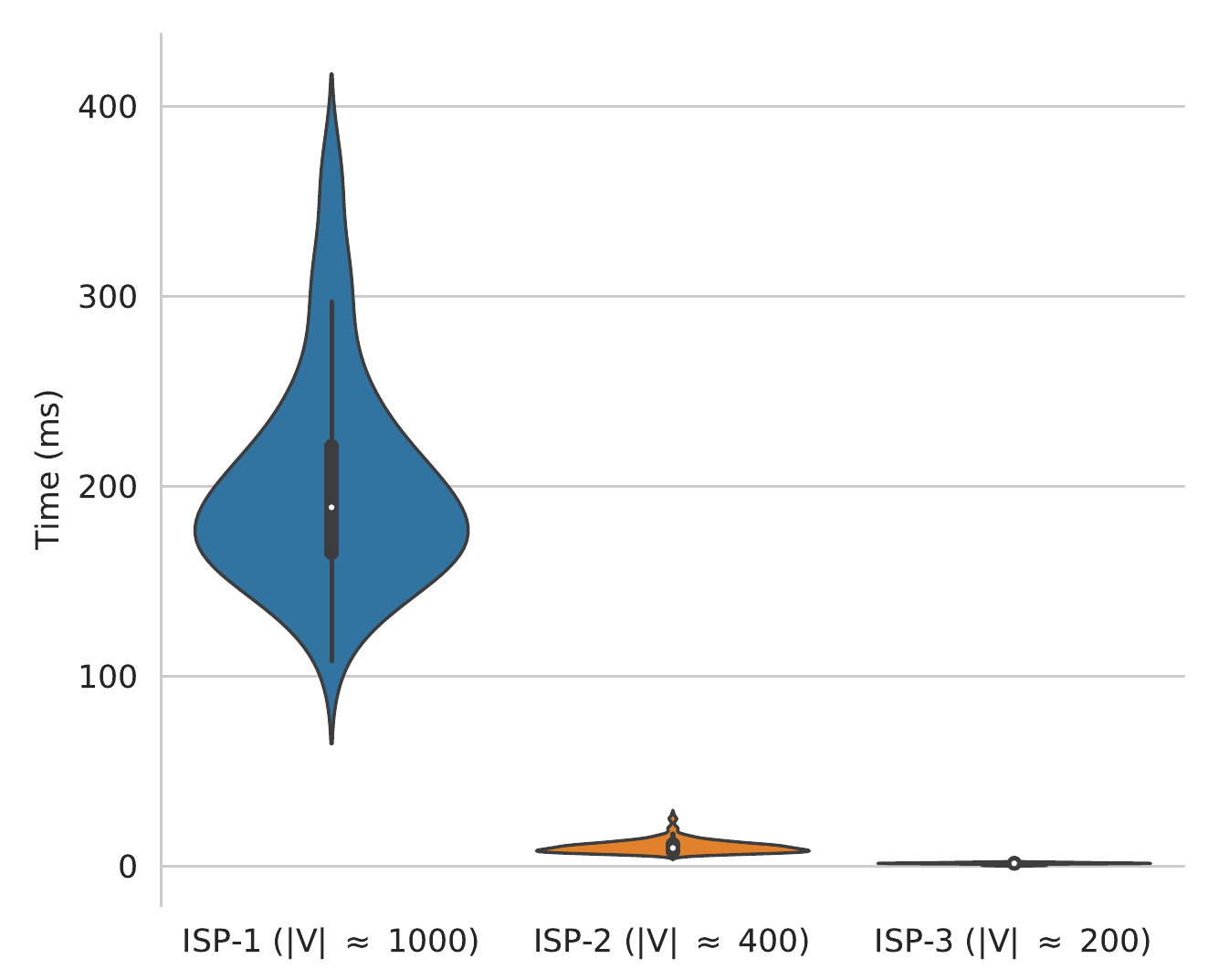}
  \caption{\bestcop's performances on realistic topologies with realistic weights (but random delays for ISP1). Execution times remain mostly under 100ms,
  even though some of the delays are still randomized for ISP1.}
  \label{fig:eval:rkt}
\end{figure}

The first ISP, ISP1, consists of more than 1100 nodes and 3000 edges.
While we do possess the IGP costs of each link in $E$, we do not have their
delays. Thus, we select random values and set them directly on $E$
(and not on $E'$ as in the previous evaluation). More precisely, we
consider here a maximum delay leading to the worst experimental computing
time, which is 70. We then also consider two other real ISP networks, ISP2 and
ISP3, with respectively around 400 and 200 nodes, having real valuations for
M$_1$ and M$_2$. The execution time results are shown in Fig.~\ref{fig:eval:rkt}
as violin plots, whose widths represent the number of executions taking the time
shown on the y-axis (in ms). We run \bestcop for all sources.

\bestcop clearly benefits from ground graph properties. Its
execution time rarely exceeds 250ms in the most disadvantageous experiments on
ISP1. For ISP2 and ISP3, the computing is almost negligible, mostly because
$|V|$ is limited. The execution times were greatly enhanced thanks to the
realistic network structures and weights leading to small Pareto fronts (few
distances dominate all the others because metrics are often aligned). Even
though the delay is still random for ISP1, simply using a realistic network
structure divided the execution time by two when compared to a randomly-weighted
full-mesh of similar size (see Fig.~\ref{fig:eval:randfm}). \bestcop already
shows great improvements regarding its execution time on ISP1, although it does
not benefit from real-life delays as in ISP2 and ISP3. In such cases with few
nodes, \bestcop solves 2COP in a negligible amount of
time. In realistic cases, it seems thus possible to increase $\Gamma$ to reach a
delay accuracy on the order of the micro-second while keeping the
execution time in the hundreds of milliseconds.

\section{Related Work}\label{sec:rl}
QoS routing and TE being popular subjects for several years, it is impossible to showcase here all past work. However, there are several extensive surveys ~\cite{survey02,survey2018,survey2010} that exhibit a lot of the solutions developed in the past decades.

Specific to DCLC,
DCUR~\cite{dcur} explores the network by extending paths either
through the least-delay or the least-cost path.
DCUR has been combined with Bellman-Ford to create DCBF~\cite{dcbf}, which guesses promising paths
through an estimated cost.
Closer to our work, Constrained Bellman-Ford~\cite{CBF94}, solves
DCLC exactly by exploring paths in a greedy fashion through a priority
queue indexed on their delay. CBF was extended in ~\cite{EBFA01}, which adds two heuristics to 
ease the problem, first by discretizing all metrics but one, second by only extending $k$ best paths.

Segment Routing also attracted a lot of interest from the research community, as 
can be seen in \cite{ventre_segment_2020}. 
While some SR-TE works are centered around the constrained paths problem~\cite{swarm, 10.1007/978-3-319-23219-5_41},
most of the work related to SR does not focus on DCLC, but rather 
bandwidth optimization ~\cite{7218434,Gay2017ExpectTU}, path encoding \cite{guedrez2016label,7417097}, or network resiliency \cite{8406885, 7524551}.
In addition, they usually rely on complex parametrizable 
techniques such as constraint programming or ILP which may lead to high computation times~\cite{ventre_segment_2020}. 
Some works do however use a construct similar to ours in order to prevent the need to perform conversions from network paths to segment lists.
\cite{Lazzeri2015EfficientLE}, in particular, proposes a multi-metric construct that does not however take advantage of dominated segments. In addition, they do not aim to solve DCLC, but simply use the construct to discover paths before sorting them lexicographically.

We propose an all-in-one solution, that solves 2COP and returns the corresponding list of segments.
Our approach is an exact algorithm with a straight-forward bounded
worst-case time complexity, with no parameters
requiring tuning. While other works solving DCLC usually detach
path computations and their deployment, we are, to the best of our knowledge,
the first ones to propose an algorithm that leverages SR deployment constraints to solve DCLC for SR.
In addition, while not all works evaluate the time complexity of their solution, or do so on limited topologies, we provide extensive evaluations on random and real large topologies, bounding the worst-case time complexity of \bestcop.

\section{Conclusion}\label{sec:conc}
While the management overhead of MPLS-based solutions leads to a TE winter in the
past decade, Segment Routing marked its rebirth. In particular, SR enables the
deployment of a practical solution to the well-known DCLC problem. In this
paper, we proposed an efficient \CR{multi-metric} SR construct onto which our algorithm,
\bestcop, iterates to solve DCLC in SR domains. Relying on adaptive simple structures, \bestcop leverages both an SR
operational constraint and the inherent limited accuracy of measured
delays. By natively encompassing such limits, we efficiently handle all scenarios. Through extensive evaluation, we showed that \bestcop performs well with both random and
realistic cases.

While we believe \bestcop is already efficient enough to be deployed, several improvements are possible to make it even
more scalable. First, its parallelizable nature and smart strategies for cache
reuse can be exploited.
 Similarly, to deal with really high trueness
requirements, more advanced, and flexible structures can be envisioned.
Finally, for large ISPs relying on subdivision in areas, partitioning DCLC into smaller sub-problems seems promising to further reduce the complexity of \bestcop.

\section*{Acknowledgements}
This work was partially supported by the French National Research Agency (ANR) project Nano-Net under contract ANR-18-CE25-0003.


\bibliographystyle{IEEEtran}
\bibliography{IEEEabrv,reference}

\end{document}